\documentclass[conference]{IEEEtran}
\IEEEoverridecommandlockouts
\usepackage{url}
\usepackage{cite}
\usepackage{amsmath,amssymb,amsfonts}
\usepackage{algorithmic}
\usepackage{graphicx}
\usepackage{textcomp}
\usepackage{xcolor}
\usepackage{NotationMacros}
\usepackage{xcolor}
\usepackage{booktabs}

\def\BibTeX{{\rm B\kern-.05em{\sc i\kern-.025em b}\kern-.08em
    T\kern-.1667em\lower.7ex\hbox{E}\kern-.125emX}}
\begin{document}

\title{FLAMO: An Open-Source Library for Frequency-Domain Differentiable Audio Processing\\
\thanks{The Aalto University School of Electrical Engineering funded the work of the first author.}
}

\author{\IEEEauthorblockN{Gloria Dal Santo$^1$, Gian Marco De Bortoli$^1$, Karolina Prawda$^2$, Sebastian J. Schlecht$^3$, Vesa Välimäki$^1$}
\IEEEauthorblockA{$^1$\textit{Acoustics Lab, Dpt. of Information and Communications Engineering,} \textit{Aalto University}, Espoo, Finland}
\IEEEauthorblockA{$^2$\textit{AudioLab, School of Physics, Engineering, and Technology,} \textit{University of York,} York, United Kingdom }
\IEEEauthorblockA{$^3$\textit{Multimedia Communications and Signal Processing, Friedrich-Alexander-Universität Erlangen-Nürnberg (FAU),} Erlangen, Germany \\
gloria.dalsanto@aalto.fi; gian.debortoli@aalto.fi; karolina.prawda@york.ac.uk; sebastian.schlecht@feu.de; vesa.valimaki@aalto.fi}}
\maketitle
\begin{abstract}
We present \textit{FLAMO}, a Frequency-sampling Library for Audio-Module Optimization designed to implement and optimize differentiable linear time-invariant audio systems. The library is open-source and built on the frequency-sampling filter design method, allowing for the creation of differentiable modules that can be used stand-alone or within the computation graph of neural networks, simplifying the development of differentiable audio systems. It includes predefined filtering modules and auxiliary classes for constructing, training, and logging the optimized systems, all accessible through an intuitive interface. Practical application of these modules is demonstrated through two case studies: the optimization of an artificial reverberator and an active acoustics system for improved response coloration.
\end{abstract}



\section{Introduction}\label{sec:intro}
As Engel's differentiable digital signal processing (DDSP)~\cite{engel2020DDSP} became a popular approach to modeling audio synthesis and processing techniques, there has been increasing interest in interpretable and differentiable audio processors such as filters, equalizers (EQs), and reverberators~\cite{hayes2024review, kuznetsov2020differentiable, steinmetz2022style, lee2022differentiable, lee2024grafx}. The appeal lies in the ability to run gradient-based optimization of their parameters, embed them in neural networks, or generate their parameters via hyperconditioning \cite{nercessian2021lightweight}. 

Frequency sampling is a technique used in DDSP, where the system's response is sampled at discrete complex frequencies. 
This approach is particularly useful for linear time-invariant (LTI) systems, as the convolution operation allows factorization of their frequency response into a cascade of processing units with multiplied frequency responses.
More often, frequency sampling is used to obtain a finite impulse response (FIR) approximation of infinite impulse response (IIR) filters~\cite{hayes2024review},  simplifying the implementation compared to time-domain methods, which suffer from vanishing/exploding gradients, high memory costs, and slow training~\cite{hayes2024review, steinmetz2022style, kuznetsov2020differentiable}. Using frequency-sampling, Nercessian~\cite{nercessian2020neural} realized a parametric EQ using differentiable biquads, later extending it to model audio distortion effects~\cite{nercessian2021lightweight}.  Similarly, Lee at al.~\cite{lee2022differentiable} applied this approach to create differentiable artificial reverberators using filtered velvet noise~\cite{valimaki2017late} and feedback delay networks (FDN)~\cite{jot1992analysis}, and used them to train a neural network as a parameter estimator. 
The authors used frequency sampling to optimize 
an FDN for smooth, natural-sounding reverbs~\cite{dalsanto2023colorlessfdn, santo2024feedback}. Similarly, the authors optimized an active acoustics (AA) system to achieve a flat feedback loop magnitude response and improve system stability~\cite{debortoli2024diffaa}.

This paper presents \textit{FLAMO}, a Python library for trainable DDSP modules based on frequency sampling. 
The library is developed using the PyTorch framework for automatic differentiation~\cite{paszke2019pytorch}. It is freely available online\footnotemark[1], under a permissive open-source license, and comes with comprehensive documentation\footnotemark[2]. We demonstrate its application in artificial reverberation and AA by replicating the author's previous work~\cite{dalsanto2023colorlessfdn, santo2024feedback, debortoli2024diffaa} using the library's classes and modules. 

One issue with frequency sampling is balancing accuracy and computational efficiency. Higher sampling densities can provide more accurate filter response approximations by reducing time-aliasing~\cite{smith2007math}, but they also increase the computational load. Moreover, ensuring the stability of IIR filters can be challenging. To address these challenges, we propose an anti-aliasing method that uses exponential decaying time envelopes and assignable mapping functions to map raw filter parameters to a target distribution. 
In addition, \textit{FLAMO} provides methods for logging intermediate results, domain flexibility, integration with neural networks, and pre-configured trainer and dataset classes, making it a versatile tool for the DDSP community.

The paper is structured as follows. Sec.~\ref{sec:method} provides background on frequency-domain optimization and the proposed time-aliasing reduction method. 
Sec.~\ref{sec:library} gives an overview of the library's core and utility classes. Sec.~\ref{sec:applications} validates the library's effectiveness. Sec.~\ref{sec:conclusions} concludes the paper.

\footnotetext[1]{{\protect{\url{https://github.com/gdalsanto/flamo}}}}
\footnotetext[2]{{\protect{\url{https://gdalsanto.github.io/flamo}}}}

\section{Frequency-domain optimization} \label{sec:method}
\subsection{Frequency sampling method}
IIR filters can be found in many audio effects, such as reverbs, EQs, and tone-shaping filters~\cite{valimaki2012fifty, valimaki2016all}.
While differentiable FIR filters can be implemented directly by
convolutional layers~\cite{hayes2024review}, implementing IIR filters is more challenging due to their recursive nature and stability requirements. 
To avoid the drawbacks of time-domain optimization~\cite{hayes2024review, steinmetz2022style, kuznetsov2020differentiable}, the frequency sampling method can be used to create an FIR approximation of IIR filters by sampling the filter's frequency response at discrete complex frequencies~\cite{nercessian2020neural, nercessian2021lightweight} whose resolution is chosen to minimize time-aliasing distortion~\cite{nercessian2021lightweight, lee2022differentiable}. 

The response of a real-valued filter is sampled on a vector of linearly-spaced frequencies from 0 to $\pi$ rad/s:
\begin{align}\label{eq:freq_vector}
    \vec{z}_{M} = [e^{\jmath \pi \frac{0}{M-1}}, e^{\jmath \pi \frac{1}{M-1}}, \dots, e^{\jmath \pi \frac{M-2}{M-1}},  e^{\jmath \pi}],
\end{align}
where $M$ is number of frequency bins, and $\jmath = \sqrt{-1}$. 
The frequency response of a single-input and single-output IIR filter $H$ of order $N$ can be evaluated over $\vec{z}_M$ as follows 
\begin{align}\label{eq:freqz}
H(\vec{z}_M) = \frac{\text{DFT}(\mathbf{b})}{\text{DFT}(\mathbf{a})} = \frac{b_0+b_1\vec{z}_M^{-1}+\dots+b_{N}\vec{z}_M^{-N+1}}{a_0+a_1\vec{z}_M^{-1}+\dots+a_{N}\vec{z}_M^{-N+1}},
\end{align}
where coefficients $\mathbf{b} = [b_0, b_1, \dots, b_{N}]$ and $\mathbf{a} = [a_0, a_1, \dots, a_{N}]$ are the feedforward and feedback gains of the filter, respectively. 
We assume real-time-domain signals such that their discrete Fourier transform (DFT) is described completely by the frequency bins in $\vec{z}_{M}$. 

The overall frequency response of two cascaded systems, denoted as $\mat{H}_1$ and $ \mat{H}_2$, is obtained by multiplying their frequency response matrices, as long as the input and output dimensions are consistent,
\begin{align}\label{eq:prod}
\mat{H}_{\textrm{series}}(\vec{z}_M) = \mat{H}_1(\vec{z}_M)\mat{H}_2(\vec{z}_M),
\end{align}
where the use of boldface notation indicates a multiple-input and multiple-output (MIMO) system.
Meanwhile, for a recursive system with feedforward path described by $\mat{G}$ and feedback path described by $\mat{F}$, the resulting frequency response is determined by
\begin{align}\label{eq:recursion}
\mat{H}_{\textrm{recursion}}(\vec{z}_M) = (\mat{I}-\mat{G}(\vec{z}_M)\mat{F}(\vec{z}_M))^{-1}\mat{G}(\vec{z}_M),
\end{align}
where $\mat{I}$ is the identity matrix of the same size as $\mat{G}\mat{F}$.

\begin{figure}[t!]
\centerline{\includegraphics[trim={1.25cm 0.2cm 1.75cm 0.5cm}, clip, width=0.5\textwidth]{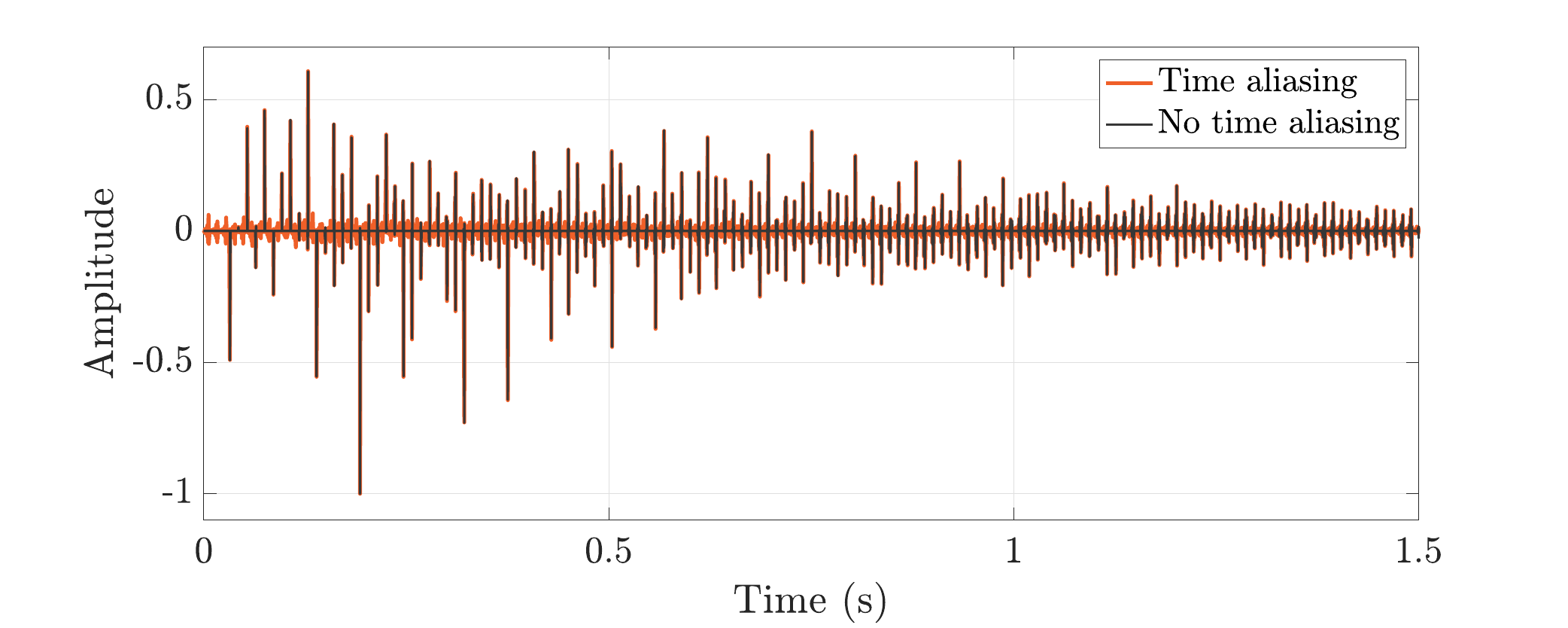}}
\caption{IR of an FDN decaying with $T_{60} = 9$\,s with (orange) and without (black) time aliasing.}
\label{fig:time_aliasing}
\end{figure}

To determine $h[n]$ we perform the inverse DFT (IDFT) by using the Hermitian property of real-valued signals on the frequency response $H(\vec{z}_M
)$. However, if the effective duration $L$ of the impulse response (IR)  exceeds the Fourier transform length $2(M-1)$, $h[n]$ will experience aliasing~\cite{smith2007math, lee2022differentiable, hayes2024review}. This issue is especially significant for lossless systems, where there are frequency points $z_m = e^{\jmath \pi\frac{m}{M-1}}$ such that $H(z_m) = \infty$. 
For lossy IIR filters, the time-aliasing error asymptotically decreases as the value of $M$ increases~\cite{lee2022differentiable}, at the cost of higher computational expense. 
Fig.~\ref{fig:time_aliasing} demonstrates time aliasing in the IR of an FDN used to synthesize reverb with homogeneous decay~\cite{jot1991digital}, characterized by a reverberation time $T_{60}$
of 9\,s. To produce the IR, we performed DFT with $M=2f_\textrm{s}$ (orange) and $M = 10 f_\textrm{s}$ (black), where $f_\textrm{s}$ is the sampling frequency. Due to time aliasing, the orange curve exhibits spurious impulses both before and after the actual onset time. 

\subsection{Time-aliasing mitigation} 
To mitigate time-aliasing, we propose an anti-aliasing method that involves applying an exponentially decaying envelope, denoted as $\gamma^n$ for $n\geq0$, before transforming to the frequency domain. 
The IR of each module is then given by 
\begin{align}\label{eq:antialias_ir}
    \hat{h}[n] = h[n] \gamma^n ,
\end{align}
where $0<\gamma\leq1$ is the decay parameter.

Using the DFT pair $f[n]\gamma^n = f[n] e^{n \ln{\gamma}}$ and $F(\omega + \jmath\ln{\gamma})$, where $f[n]$ is a real function with DFT $F(\omega)$ and $\omega$ is the angular frequency in rad/s,  we can express the frequency response of~\eqref{eq:antialias_ir} as $\hat{H}(\ejw) = H(e^{\jmath (\omega + \jmath\ln{\gamma})}) = H( e^{\jmath \omega}/ \gamma)$.
Thus, introducing an exponential decay in the time domain corresponds to sampling the frequency response outside the unit circle, with a radius of $1 / \gamma$. The original IR $h$ can then be recovered from $\hat{h}$ by compensating for the decay in time-domain, i.e.,
\begin{align}
\begin{aligned}
    h[n] = \hat{h}[n] \gamma^{-n} = \mathrm{IDFT}(\hat{H}(\ejw)) \gamma^{-n}.
\label{eq:antialiased_ir}    
\end{aligned}
\end{align}
If samples in the wrapping region during convolution are not sufficiently attenuated by $\gamma^n$, they appear at the start of the impulse response. Conversely, a very small $\gamma^n$ can amplify numerical noise, which depends on machine precision. Thus, $\gamma$ must be carefully selected to balance attenuation and noise.

The technique can be applied to a series of multiple LTI modules. By consistently applying the same exponential decay to each time-domain filter before performing the DFT, one can effectively manage time-domain aliasing. This method is particularly useful for enhancing the early response of a system.  Without such a technique, avoiding aliasing would either necessitate excessive damping of the system or require computing the entire response, both of which are inefficient.

\section{\textit{FLAMO} Library Structure} \label{sec:library}
\begin{figure}[t!]
\centerline{\includegraphics[trim={0.6cm 0cm 0.75cm 0cm},width=0.5\textwidth]{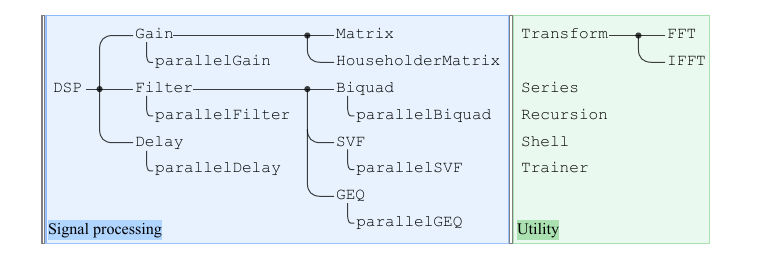}}
\vspace{-4mm}
\caption{Inheritance tree of the classes in the \textit{FLAMO} library designed for signal processing (left) and utility (right) purposes.}
\label{fig:library_overview}
\end{figure}

Fig.~\ref{fig:library_overview} provides an overview of the \textit{FLAMO} library's structure. The primary class is \texttt{DSP}, from which the \texttt{Gain}, \texttt{Filter}, and \texttt{Delay} classes inherit attributes and methods. The library's interface has been simplified by the \texttt{Trainer} and \texttt{Shell} utility classes, which allow users to set training parameters, initiate optimization, and log intermediate results. Each processing module receives the frequency response of the incoming signal as input and multiplies it by the frequency response of the system that the module represents. 
This operation is defined using notation based on the Einstein summation convention using the \texttt{torch.einsum()} function~\cite{pythonEinsum}. Frequency responses are complex-valued tensors, and we utilize PyTorch's ability to backpropagate through real-valued functions of complex tensors via Wirtinger calculus~\cite{pythonComplex}. Safety checks are implemented to ensure coherent tensor shapes and frequency-sampling parameter values.

\vspace{-0.25mm}

Each module operates as a MIMO system. In the case of parallel processing, where each channel of the incoming signal passes through a different instance of the same system, element-wise vector multiplication of the frequency responses can be used to reduce the number of computations. For this purpose, each class has a corresponding \texttt{parallel*} version, where \texttt{*} denotes the original system module. A summary of the shapes of the tensors expected and produced by each class is provided in Table~\ref{tb:tensor_shapes}, from which the \texttt{Delay} class has been omitted since it shares the same shapes as the \texttt{Filter} class. 

\subsection{Digital signal processor class}
The \texttt{DSP} class represents a generic learnable module within the library. It inherits from PyTorch's \texttt{nn.Module}~\cite{pythonModule}, with its primary attribute, \texttt{param}, representing the learnable parameters for each instance of the class. These parameters can be mapped from the initial distribution to a user-specified distribution via the \texttt{map} argument. By default, the distribution is set to normal at initialization. Some inherited classes built into the library have predefined mappings to ensure system stability or to achieve filter-specific parameterizations.

The \texttt{Gain} and  \texttt{Matrix} child classes implement systems of broadband gains. The latter allows one to choose among a set of pre-built mappings, such as mappings to orthogonal matrices~\cite{lezcano2019cheap}. For FIR filters one can use the \texttt{Filter} class.

The library provides subclasses for the most commonly used filter parameterizations, including biquad, state variable filter (SVF)~\cite{wishnick2014time, kuznetsov2020differentiable}, and graphic EQs (GEQs)~\cite{valimaki2016all, schlecht2017accurate}. The \texttt{Biquad} class implements low-pass, high-pass, and band-pass filters using RBJ cookbook formulas~\cite{bristow2024cookbook}, which map the cutoff frequency and gain parameters to the $\mathbf{b}$ and $\mathbf{a}$ coefficients. For higher-order filters, the \texttt{SVF} class allows cascading multiple biquad filters parameterized by SVF parameters. These parameters offer simpler stability conditions~\cite{wishnick2014time, kuznetsov2020differentiable} and often outperform raw biquad parameters~\cite{lee2022differentiable}. The \texttt{SVF} class includes pre-built mappings for second-order low-pass, high-pass, and band-pass filters, as well as low-shelf, high-shelf, peaking, and notch filters~\cite{wishnick2014time}. Alternatively, the class can initialize a generic SVF filter with random mixing coefficients. The \texttt{GEQ} class implements GEQs as described in~\cite{schlecht2017accurate} for full and one-third octave intervals.
 
\begin{table}[t!]
\caption{Frequency response shapes for each module, their expected input, and the produced output. $N_\textrm{in}\,\text{and}\,N_\textrm{out}$ are the input and output dimensions, $M$ is the number of frequency points, and $B$ is the batch size.}\label{tb:tensor_shapes}
\vspace{-0.25mm}
\begin{tabular*}{0.5\textwidth}{@{\extracolsep\fill}lrrr}
\toprule 
 Class & Module $\mat{H}(z)$ & Input & Output \\ %
\midrule
 \texttt{Gain} & $N_\textrm{out} N_\textrm{in}$ & $B  M  N_\textrm{in}$ & $B  M  N_\textrm{out}$ \\
 \texttt{parallelGain} & $ N_\textrm{in}$  & $B  M  N_\textrm{in}$ & $B  M  N_\textrm{in}$\\
 \texttt{Filter} & $M  N_\textrm{out} N_\textrm{in}$ & $B  M  N_\textrm{in}$ & $B  M  N_\textrm{out}$ \\
 \texttt{parallelFilter} & $ M  N_\textrm{in}$  & $B  M  N_\textrm{in}$ & $B  M  N_\textrm{in}$\\
\midrule
\end{tabular*}
\end{table} 

Differentiable delays are implemented directly in the frequency domain via the complex exponential $\vec{z}_M^{-m}$, where $m$ is the delay in samples. By default, delays are fractional, but integer sample delays can be requested. The learnable parameter of the \texttt{Delay} class is the delay length in seconds. We recommend adjusting the \texttt{unit} attribute, which controls the time unit, as using subdivisions or multiples of time can improve update effectiveness in certain applications.
\subsection{Utility classes}
The \texttt{Series} class extends PyTorch's \texttt{nn.Sequential}~\cite{pythonSequential} to maintain coherence among chained modules and prevent error-prone nesting. 
The \texttt{Recursion} class implements a recursive closed-loop frequency response~\eqref{eq:recursion} having processing modules as feedforward and feedback paths. 
The \texttt{Shell} class is a container designed to interface the user-defined DSP chain with the desired loss function and to provide auxiliary functionalities. It has three main attributes: \texttt{core}, \texttt{input\textunderscore layer}, and \texttt{output\textunderscore layer}. The \texttt{core} represents the differentiable system to be optimized, while \texttt{input\textunderscore layer} and \texttt{output\textunderscore layer} handle transformations between the system, the dataset input, and the loss function. The \texttt{Shell} keeps these components separate, allowing the system to be defined independently of the loss function and dataset. Additionally, the class includes methods to return the system's time and frequency responses by temporarily modifying the input and output layers.

\texttt{Series}, \texttt{Recursion}, and \texttt{Shell} classes also ensure that the operational flow attributes, e.g. $M$, are consistent across all modules, between feedforward and feedback paths, and between \texttt{input\textunderscore layer}, \texttt{core}, and \texttt{output\textunderscore layer}, respectively. The \texttt{Trainer} class manages the optimization of the modules within a system, it handles key functionalities such as initialization of training parameters, criterion management, training loop, and logging of intermediate states.  
\section{Applications}
\label{sec:applications}
\begin{figure}[t!]
\centerline{\includegraphics[trim={2.3cm 0.5cm 1.2cm 1.5cm},width=0.45\textwidth]{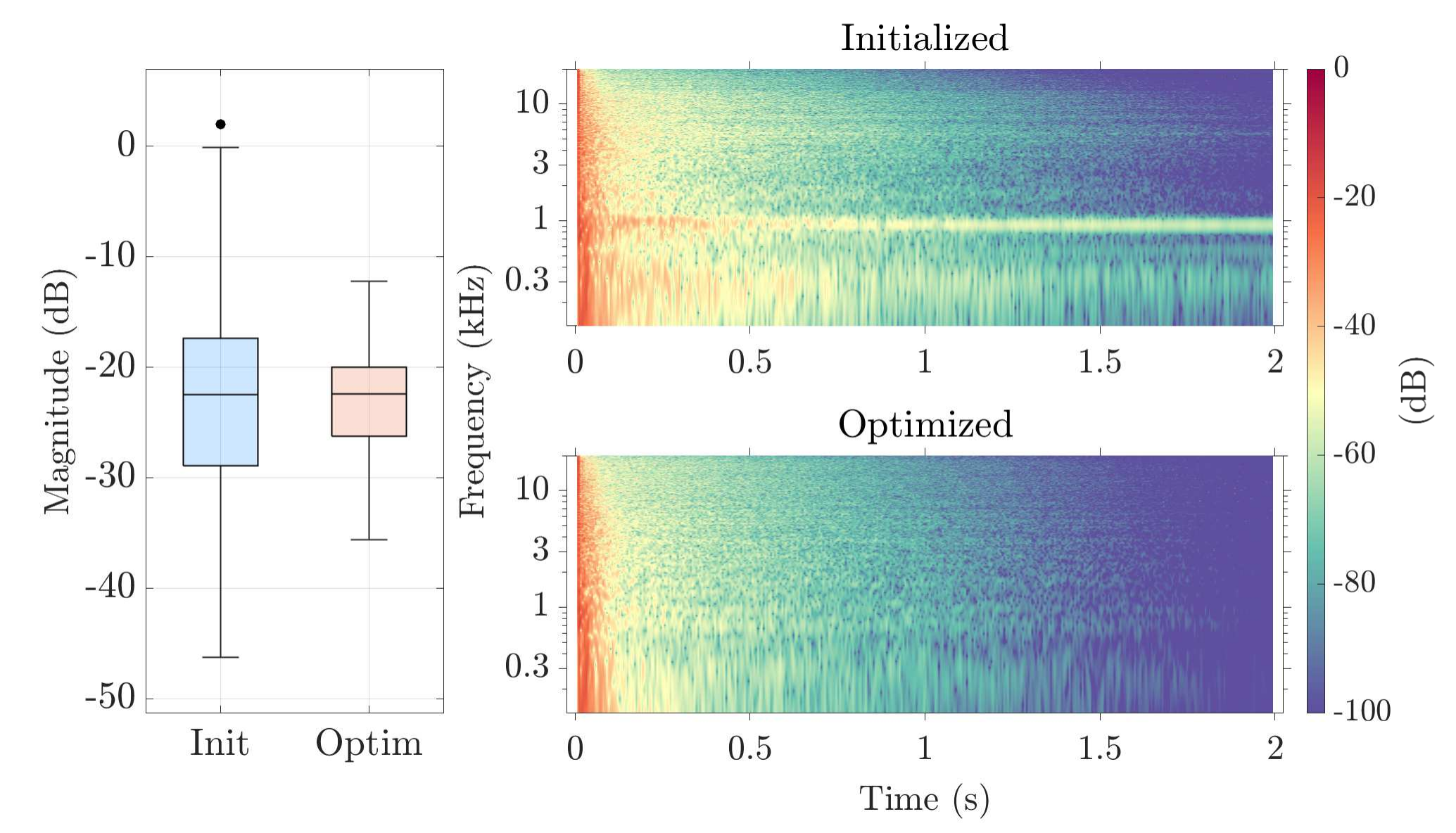}}
\vspace{-2mm}
\caption{(Left) Eigenvalue magnitude distribution and (right) spectrogram of 
an active acoustics system before and after optimization for spectral flatness.}
\label{fig:aa_optim}
\end{figure}
\subsection{Active acoustics}
AA systems are used to electronically provide controlled variability in the acoustics of closed spaces~\cite{svensson1996reverberation, poletti2000stability, poletti2011active}. They consist of microphones, loudspeakers, and a DSP unit. 
The AA system must be well-tuned to blend the artificial sound field with the natural-room sound field and to avoid coloration artifacts. The tuning process requires time and expertise, as many parameters, e.g. transducer gains and filter coefficients, must be carefully chosen. Previously, the authors have demonstrated the applicability of DDSP in the AA scenario to automatize the tuning process to flatten the spectral behavior, which led to a reduced coloration in the full-system IR~\cite{debortoli2024diffaa}.

We redefine the former design~\cite{debortoli2024diffaa} with the \textit{FLAMO} library. The \texttt{Filter} class implements the learnable FIR matrix $\mat{U}(z)$ and a fixed white-Gaussian-noise-reverb matrix $\mat{R}(z)$. 
The general system gain $\mat{G}$ is built with the \texttt{parallelGain} class. 
The loudspeaker-to-microphone IR matrix $\mat{H}_\mathrm{LM}$ is implemented as a \texttt{Filter}-class instance of the measured room IRs. 
The transfer function matrix describing one feedback-loop iteration is $\mat{F}_\mathrm{MM}(z) = \mat{G} \mat{H}_\mathrm{LM}(z) \mat{R}(z) \mat{U}(z)$.
We run optimization to flatten the frequency response of $\mat{F}_\mathrm{MM}(z)$ and improve the system stability. Fig.~\ref{fig:aa_optim} presents the optimization results for the magnitude eigenvalue distribution and the spectrograms of the full-system IR. In the box plot, the lower and upper edges represent the 25th and 75th percentiles, respectively, highlighting the narrowing of the magnitude eigenvalue distribution from initialization. The coloration artifacts, indicated by long-ringing modes in the spectrograms, are also considerably attenuated. The black dot in Fig.~\ref{fig:aa_optim} indicates the maximum outlier of the
initial and optimized eigenvalue magnitude distribution and it is responsible for the strong resonance at 1 kHz in the
top-spectrogram. Further outliers have not been considered as they don’t add information regarding system stability or spectral flatness.


\subsection{Delay-line-based artificial reverberation}
\begin{figure}[t!]
\vspace{-3mm}
\centerline{\includegraphics[trim={1.2cm 0.2cm 1cm 1.5
cm},width=0.44\textwidth]{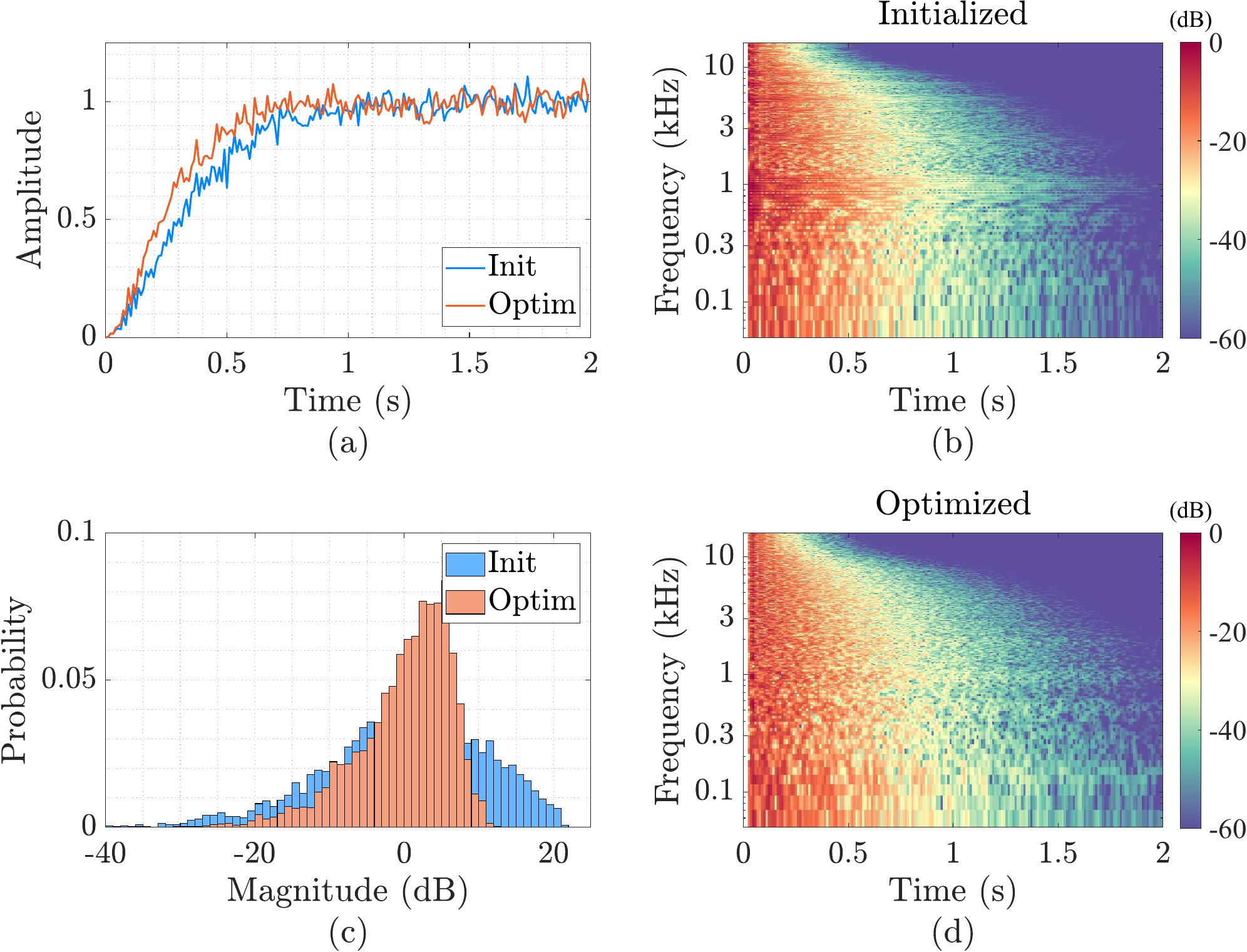}}
\vspace{-3mm}
\caption{Effect of the optimization of an FDN of size $N=6$ with gain coefficient optimized to improve smoothness. Echo density profile (a) modal excitation distribution (c) and spectrogram of the IR (b) before and (d) after optimization.} 
\label{fig:fdn_optim}
\end{figure}
The FDN is a recursive system used in reverb synthesis~\cite{jot1991digital, jot1992analysis, valimaki2012fifty} and consists of delay lines $\mat{m}$, a set of input and output gains $\mat{b}$ and $\mat{c}$, and a scalar feedback matrix through which the delay outputs are coupled to the delay inputs.
The transfer function of the FDN is 
\begin{align}\label{eq:tr_fdn}
    H(z) = \mat{c}^\top\big[\mat{D_m}(z)^{-1} -\mat{A}\big]^{-1}\mat{b} + d\, 
\end{align}
where $\mat{A}$ is the $N\times N$ feedback matrix, with $N$ being the number of delay lines, $\mat{D_m}(z)$ is the diagonal delay matrix, and $d$ is the direct gain. The operator $(\cdot)^\top$ denotes the transpose. A common design choice is to start with a lossless FDN prototype with a smooth response. This can be achieved by using an orthogonal matrix $\mat{A}$~\cite{schlecht2016feedback, stautner1982designing} optimized for spectral flatness and temporal density~\cite{dalsanto2023colorlessfdn, santo2024feedback}. Attenuation filters are then inserted to control frequency-dependent $T_{60}$~\cite{jot1992analysis, Valimaki2024, schlecht2017accurate, prawda2019improved}. 
We implement and optimize a lossless FDN with temporal anti-aliasing using \textit{FLAMO}. This approach contrasts with \cite{dalsanto2023colorlessfdn, santo2024feedback}, where the system had to be made lossy to enable backpropagation. 
Matrix $\mat{A}$ is implemented using a learnable \texttt{Matrix}-class instance with orthogonal mapping. Input and output gains are learnable \texttt{Gain}-class instances. Delays of given lengths are an instance of the \texttt{parallelDelay} class. The attenuation filters are then implemented using the class \texttt{parallelGEQ}, with the parameter mapping proposed in~\cite{schlecht2017accurate}, and inserted in the optimized system. 

Improvements in temporal density can be observed from the echo density profile~\cite{abel2006simple}, which is representative of the distribution of reflections over time. Fig.~\ref{fig:fdn_optim}~(a) shows a faster echo build-up achieved by the optimized FDN. Reduction in coloration is reflected by a narrower distribution of the modal excitation~\cite{heldmann2021role} depicted in Fig.~\ref{fig:fdn_optim}~(c), where a significant attenuation of the loudest modes can be observed.
Fig.~\ref{fig:fdn_optim}~(b) and~(d) show the spectrogram of the system's response before and after optimization, respectively. The optimization results in more evenly distributed energy across the frequency range, reducing coloration artifacts.

\section{Conclusion}\label{sec:conclusions}
The \textit{FLAMO} library for frequency-sampling optimization of differentiable audio modules 
simplifies the implementation and optimization of differentiable LTI systems, offering an intuitive API, and can be integrated with PyTorch neural networks. 
We validate the library's effectiveness through artificial and enhanced reverberation scenarios. The source code is available in an online repository\footnotemark[1] and on the Python package index\footnotemark[3]. We plan to continuously expand the \textit{FLAMO} library to support the audio signal processing community.
\footnotetext[3]{{\protect{\url{https://pypi.org/project/flamo}}}}

\bibliographystyle{ieeetr}
\bibliography{bibliography}

\begin{thebibliography}{10}

\bibitem{engel2020DDSP}
J.~Engel, L.~H. Hantrakul, C.~Gu, and A.~Roberts, ``{DDSP}: Differentiable digital signal processing,'' in {\em Proc. Int. Conf. Learning Representations}, 2020.

\bibitem{hayes2024review}
B.~Hayes, J.~Shier, G.~Fazekas, A.~McPherson, and C.~Saitis, ``A review of differentiable digital signal processing for music and speech synthesis,'' {\em Frontiers in Signal Processing}, vol.~3, 2024.

\bibitem{kuznetsov2020differentiable}
B.~Kuznetsov, J.~D. Parker, and F.~Esqueda, ``Differentiable {IIR} filters for machine learning applications,'' in {\em Proc. 23rd Int. Conf. Digital Audio Effects}, pp.~297--303, 2020.

\bibitem{steinmetz2022style}
C.~J. Steinmetz, N.~J. Bryan, and J.~D. Reiss, ``Style transfer of audio effects with differentiable signal processing,'' {\em J. Audio Eng. Soc.}, vol.~70, pp.~708--721, 2022.

\bibitem{lee2022differentiable}
S.~Lee, H.-S. Choi, and K.~Lee, ``Differentiable artificial reverberation,'' {\em IEEE/ACM Trans. Audio, Speech, Lang. Process.}, vol.~30, pp.~2541--2556, 2022.

\bibitem{lee2024grafx}
S.~Lee, M.~Mart{\'\i}nez-Ram{\'\i}rez, W.-H. Liao, S.~Uhlich, G.~Fabbro, K.~Lee, and Y.~Mitsufuji, ``Grafx: an open-source library for audio processing graphs in pytorch,'' {\em arXiv preprint arXiv:2408.03204}, 2024.

\bibitem{nercessian2021lightweight}
S.~Nercessian, A.~Sarroff, and K.~J. Werner, ``Lightweight and interpretable neural modeling of an audio distortion effect using hyperconditioned differentiable biquads,'' in {\em IEEE Int. Conf. Acoust., Speech, Signal Process. (ICASSP)}, pp.~890--894, IEEE, 2021.

\bibitem{nercessian2020neural}
S.~Nercessian, ``Neural parametric equalizer matching using differentiable biquads,'' in {\em Proc. 23rd Int. Conf. Digital Audio Effects}, pp.~265--272, 2020.

\bibitem{valimaki2017late}
V.~V{\"a}lim{\"a}ki, B.~Holm-Rasmussen, B.~Alary, and H.-M. Lehtonen, ``Late reverberation synthesis using filtered velvet noise,'' {\em Appl. Sci.}, vol.~7, no.~5, p.~483, 2017.

\bibitem{jot1992analysis}
J.-M. Jot, ``An analysis/synthesis approach to real-time artificial reverberation,'' in {\em Proc. IEEE Int. Conf. Acoust. Speech Signal Process. (ICASSP)}, vol.~2, pp.~221--224, 1992.

\bibitem{dalsanto2023colorlessfdn}
G.~{Dal Santo}, K.~Prawda, S.~J. Schlecht, and V.~V{\"a}lim{\"a}ki, ``Differentiable feedback delay network for colorless reverberation,'' in {\em Proc. 26th Int. Conf. Digital Audio Effects}, (Copenhagen, Denmark), pp.~244--251, 2023.

\bibitem{santo2024feedback}
G.~{Dal Santo}, K.~Prawda, S.~J. Schlecht, and V.~V{\"a}lim{\"a}ki, ``Optimizing tiny colorless feedback delay networks,'' {\em arXiv preprint arXiv:2402.11216}, 2024.

\bibitem{debortoli2024diffaa}
G.~M. {De Bortoli}, G.~{Dal Santo}, K.~Prawda, V.~V{\"a}lim{\"a}ki, T.~Lokki, and S.~J. Schlecht, ``Differentiable active acoustics,'' in {\em Proc. 27th Int. Conf. Digital Audio Effects}, (Guildford, UK), 2024.

\bibitem{paszke2019pytorch}
A.~Paszke, S.~Gross, F.~Massa, A.~Lerer, J.~Bradbury, G.~Chanan, T.~Killeen, Z.~Lin, N.~Gimelshein, L.~Antiga, {\em et~al.}, ``Pytorch: An imperative style, high-performance deep learning library,'' {\em Advances in neural information processing systems}, vol.~32, 2019.

\bibitem{smith2007math}
J.~O. Smith, {\em Mathematics of the Discrete Fourier Transform (DFT)}.
\newblock W3K, 2007.

\bibitem{valimaki2012fifty}
V.~Välimäki, J.~D. Parker, L.~Savioja, J.~O. Smith, and J.~S. Abel, ``Fifty years of artificial reverberation,'' {\em IEEE Trans. Audio, Speech, Lang. Process.}, vol.~20, no.~5, pp.~1421--1448, 2012.

\bibitem{valimaki2016all}
V.~V{\"a}lim{\"a}ki and J.~D. Reiss, ``All about audio equalization: Solutions and frontiers,'' {\em Appl. Sci.}, vol.~6, no.~5, p.~129, 2016.

\bibitem{jot1991digital}
J.-M. Jot and A.~Chaigne, ``Digital delay networks for designing artificial reverberators,'' in {\em Audio Engineering Society Convention 90}, 1991.

\bibitem{pythonEinsum}
``{PyTorch} torch.einsum.'' \url{https://pytorch.org/docs/stable/generated/torch.einsum.html}.
\newblock Accessed: 2024-09-8.

\bibitem{pythonComplex}
``{PyTorch} complex numbers.'' \url{https://pytorch.org/docs/stable/complex_numbers.html}.
\newblock Accessed: 2024-08-23.

\bibitem{pythonModule}
``{PyTorch} module.'' \url{https://pytorch.org/docs/stable/generated/torch.nn.Module.html}.
\newblock Accessed: 2024-09-8.

\bibitem{lezcano2019cheap}
M.~Lezcano-Casado and D.~Mart{\i}nez-Rubio, ``Cheap orthogonal constraints in neural networks: A simple parametrization of the orthogonal and unitary group,'' in {\em Proc. Int. Conf. Machine Learning}, pp.~3794--3803, 2019.

\bibitem{wishnick2014time}
A.~Wishnick, ``Time-varying filters for musical applications.,'' in {\em Proc. 17th Int. Conf. Digital Audio Effects}, pp.~69--76, 2014.

\bibitem{schlecht2017accurate}
S.~J. Schlecht and E.~A. Habets, ``Accurate reverberation time control in feedback delay networks,'' in {\em Proc. 20th Int. Conf. Digital Audio Effects, Edinburgh, UK}, pp.~337--344, 2017.

\bibitem{bristow2024cookbook}
R.~Bristow-Johnson, ``{RBJ Audio EQ Cookbook}.'' \url{https://webaudio.github.io/Audio-EQ-Cookbook/Audio-EQ-Cookbook.txt}, 2005.
\newblock [Online; accessed 19-August-2024].

\bibitem{pythonSequential}
``{PyTorch} sequential.'' \url{https://pytorch.org/docs/stable/generated/torch.nn.Sequential.html}.
\newblock Accessed: 2024-09-8.

\bibitem{svensson1996reverberation}
P.~U. Svensson, {\em On Reverberation Enhancement in Auditoria}.
\newblock PhD thesis, Chalmers University of Technology, G{\"o}teborg, Sweden, 1994.

\bibitem{poletti2000stability}
M.~Poletti, ``The stability of single and multichannel sound systems,'' {\em Acta Acust. un. Acust.}, vol.~86, no.~1, pp.~163--178, 2000.

\bibitem{poletti2011active}
M.~Poletti, ``Active acoustic systems for the control of room acoustics,'' {\em Build. Acoust.}, vol.~18, no.~3--4, pp.~237--258, 2011.

\bibitem{schlecht2016feedback}
S.~J. Schlecht and E.~A. Habets, ``Feedback delay networks: Echo density and mixing time,'' {\em IEEE/ACM Trans. Audio, Speech, Lang. Process.}, vol.~25, no.~2, pp.~374--383, 2016.

\bibitem{stautner1982designing}
J.~Stautner and M.~Puckette, ``Designing multi-channel reverberators,'' {\em Computer Music Journal}, vol.~6, no.~1, pp.~52--65, 1982.

\bibitem{Valimaki2024}
V.~Välimäki, K.~Prawda, and S.~J. Schlecht, ``Two-stage attenuation filter for artificial reverberation,'' {\em IEEE Signal Process. Lett.}, vol.~31, pp.~391--395, 2024.

\bibitem{prawda2019improved}
K.~Prawda, S.~J. Schlecht, and V.~V{\"a}lim{\"a}ki, ``Improved reverberation time control for feedback delay networks,'' in {\em Proc. 22nd Int. Conf. Digital Audio Effects}, pp.~1--7, 2019.

\bibitem{abel2006simple}
J.~S. Abel and P.~Huang, ``A simple, robust measure of reverberation echo density,'' in {\em Proceedings of the Audio Engineering Society Convention 121}, Audio Engineering Society, 2006.

\bibitem{heldmann2021role}
J.~Heldmann and S.~J. Schlecht, ``The role of modal excitation in colorless reverberation,'' in {\em Proc. 24th Int. Conf. Digital Audio Effects}, pp.~206--213, 2021.

\end{thebibliography}

\end{document}